\documentclass[a4paper,11pt]{article}
\pdfoutput=1 

\usepackage{jcappub} 

\usepackage[T1]{fontenc} 
\usepackage{comment}
\usepackage{braket}
\usepackage{amsmath}
\usepackage{soul}
\usepackage{makecell}
\usepackage{verbatim,acronym}
\usepackage{subfigure}
\usepackage{endnotes}
\usepackage[normalem]{ulem} 
\usepackage{epsfig}
\usepackage{slashed,skak}

\newcommand{\gev}{\, {\rm GeV}}

\newcommand{\beq}{\begin{equation}}
\newcommand{\eeq}{\end{equation}}
\newcommand{\bea}{\begin{eqnarray}}
\newcommand{\eea}{\end{eqnarray}}

\newcommand{\gsim}{\lower.7ex\hbox{$\;\stackrel{\textstyle>}{\sim}\;$}}
\newcommand{\lsim}{\lower.7ex\hbox{$\;\stackrel{\textstyle<}{\sim}\;$}}

\newcommand{\be}{\begin{equation}}
\newcommand{\ee}{\end{equation}}
\newcommand{\ba}{\begin{eqnarray}}
\newcommand{\ea}{\end{eqnarray}}

\newcommand{\D}{\mathrm{d}}

\title{\boldmath Dark Matter-Induced Low-Mass Gap Black Hole Echoing LVK Observations}

\author[a,b,c]{Shuailiang Ge}
\author[d,e]{Yuxin Liu}
\author[b,c,f]{Jing Shu}
\author[g]{and Yue Zhao}

\affiliation[a]{Department of Physics, Korea Advanced Institute of Science and Technology (KAIST), Daejeon 34141, South Korea}

\affiliation[b]{School of Physics and State Key Laboratory of Nuclear Physics and Technology, Peking University, Beijing 100871, China}
\affiliation[c]{Center for High Energy Physics, Peking University, Beijing 100871, China}
\affiliation[d]{International Centre for Theoretical Physics Asia-Pacific, Beijing/Hangzhou, 100080, China}
\affiliation[e]{University of Chinese Academy of Sciences, Beijing 100190, China}
\affiliation[f]{Beijing Laser Acceleration Innovation Center, Huairou, Beijing, 101400, China}
\affiliation[g]{Department of Physics and Astronomy, University of Utah, Salt Lake City, UT 84112, USA}

\emailAdd{sge@pku.edu.cn}
\emailAdd{liuyuxin211@mails.ucas.ac.cn}
\emailAdd{jshu@pku.edu.cn}
\emailAdd{zhaoyue@physics.utah.edu}

\abstract{The recent detection of gravitational waves from a binary merger involving a potential low-mass gap black hole (LMBH) by LIGO-Virgo-KAGRA (LVK) Collaboration motivates investigations into mechanisms beyond conventional stellar evolution theories to account for their existence. We study a mechanism in which dark matter (DM), through its capture and accumulation inside main sequence stars, induces the formation of black holes within the mass range of $[3, 5]M_\odot$. We examine the distribution of these LMBHs as a function of galaxy halo mass, particularly when paired with neutron stars. This gives a distinct signature that can be tested with future gravitational wave observations. We find that a viable portion of the DM parameter space predicts a merger rate of such binaries consistent with LVK observations.}

\begin{document}
\maketitle
\flushbottom

\section{Introduction}
The evolution of a main sequence (MS) star is complicated and varies depending on its mass. Stars within the mass range of $[0.5,8]M_\odot$, including the Sun, typically evolve into red giants after depleting their hydrogen fuel and subsequently transform into white dwarfs (WD). MS stars with masses within $[8,20]M_\odot$ possess sufficient energy to trigger supernova explosions, resulting in the violent loss of mass. This process leads to the formation of a neutron star (NS). According to the Equation of State, the maximum mass of a NS cannot exceed $3M_\odot$ ~\cite{Mueller:1996pm,kalogera1996maximum,Godzieba:2020tjn}. When the mass of the progenitor star exceeds approximately $\sim 20 M_\odot$\cite{Belczynski:2011bn}, the supernova explosion can be suppressed, leaving enough matter behind for the formation of an astrophysical black hole (ABH). The mass of the ABH directly produced through this process is unlikely to be smaller than $5M_\odot$ ~\cite{Belczynski:2011bn,OConnor:2010moj,Fryer:2011cx, Janka:2012wk, Muller:2016ujh,Ertl:2019zks}. \footnote{There are still debates regarding the validity of this statement, particularly concerning the fallback of outgoing matter from supernovae, as discussed in ~\cite{Mandel:2020qwb, Antoniadis:2021dhe, Sukhbold:2015wba, Ertl:2019zks}.}
As a consequence, there is a gap in the mass distribution of compact objects between $[3,5]M_\odot$, a feature consistent with observations of Galactic X-ray binaries ~\cite{Bailyn:1997xt,Ozel:2010su,Farr:2010tu,Kreidberg:2012ud}.

Using traditional observation methods, searching for a low-mass gap black hole (LMBH), with mass $[3,5]M_\odot$  can be challenging due to its small size and very low luminosity in electromagnetic radiation. However, gravitational wave (GW) radiation offers a new avenue for detection, even for objects located very far away from our Milky Way galaxy. Recently, the LIGO-Virgo-KAGRA (LVK) Collaboration announced the detection of a merger (GW230529)
between a NS and an LMBH with a mass of $3.6^{+0.8}_{-1.2}M_\odot$~\cite{LIGOScientific:2024elc}. 
This breakthrough opens up a new window for investigating the existence and properties of LMBHs. Particularly, the presence of a LMBH within the mass range of $[3,5] M_\odot$ may suggest phenomena beyond our current understanding, potentially necessitating novel mechanisms for their generation. For instance, such LMBHs may originate from former triple or quadruple systems ~\cite{Fragione:2020aki, Lu:2020gfh, Liu:2020gif,Vynatheya:2021mgl, Gayathri:2023met,Bartos:2023lfu}, or through dynamical capture in star clusters ~\cite{Clausen:2012zu, Gupta:2019nwj, Ye:2019xvf,Rastello:2020sru, Sedda:2020wzl, ArcaSedda:2021zmm}. Additionally, LMBHs may be identified as primordial black holes ~\cite{Bird:2016dcv,Clesse:2016vqa,Sasaki:2016jop,Kashlinsky:2016sdv,Clesse:2020ghq}. In this paper, we take the detection of the LMBH-NS merger as the motivation and study a novel mechanism for the LMBH production, through the dark matter (DM) capture.

The existence of DM is widely accepted, yet its properties remain in mystery. Numerous efforts are dedicated to studying the interaction between DM particles and ordinary matter, such as nucleons and electrons, especially through DM direct detection experiments ~\cite{CDMS:2002moo,XENON:2017vdw,PandaX-II:2017hlx,Kavanagh:2017cru,Digman:2019wdm,Clark:2020mna,Carney:2022gse}. The general interpretation of the null results in these experiments suggests a very weak interaction. On the other hand, if the interaction between DM particles and ordinary matter is too strong, the DM may not be able to freely penetrate the atmosphere and reach the experimental devices as expected. This leads to the untested extreme of DM particles with strong interactions, not explored by these experiments. 

In this study, we consider the DM mass $m_\chi$ and cross section $\sigma_{\chi\mathrm{H}}$ between dark matter and protons in the following range,
\begin{equation}\label{eq:DM_parameter_space}
    m_\chi \in [10^4, 10^9] ~\mathrm{GeV}
    ,~~~
    \sigma_{\chi\mathrm{H}} \in [10^{-27},10^{-22}]~\mathrm{cm^2}.
\end{equation} 
This overlaps with the parameter space for strongly interacting DM models which are still consistent with various experimental and experimental constraints ~\cite{Kavanagh:2017cru,Albuquerque:2003ei,Bhoonah:2020fys}.

Although DM particles in this strongly coupled regime may evade detection in terrestrial experiments, they could readily be captured and accumulate inside an MS star. The continuous accumulation of DM particles may lead to the formation of a mini BH at the stellar center, potentially altering the ultimate fate of the star. Specifically, the presence of such a mini BH may cause a star that would otherwise become a WD or a NS to instead become a LMBH.

In this paper, we first examine the criteria for efficient DM capture and subsequent collapse to form a small BH capable of surviving Hawking radiation. We then explore how the presence of such a mini BH alters the final state of a star across various mass regimes and calculate the probability of a star being converted to a LMBH within a given halo mass. We estimate the probability distribution of detecting a LMBH-NS merger as a function of halo mass. This is a unique prediction of the DM-induced LMBH formation mechanism. With the future expansion of the GW network and enhancements in GW detector sensitivities, such a distribution serves as a discriminator to differentiate this mechanism from others. At last, we show that a substantial portion of parameter space in our DM model is capable of yielding a merger rate of LMBH-NS binaries consistent with the numbers reported by LVK.

\section{Dark matter collapse into a black hole inside a star}

As a DM particle passes through a star's interior, it interacts with the stellar material, resulting in energy loss, which may cause it to be trapped inside by the star. A detailed study of such a capture process can be found in ~\cite{Gould:1987ir,Jungman:1995df,Garani:2017jcj,Acevedo:2020gro,Ray:2023auh,Leane:2023woh,Bhattacharya:2024pmp}. The capture probability is $F_{\rm cap}$, multiplying which with the DM flux hitting the star gives the DM accumulation rate within a star~\cite{Gould:1987ir,Ray:2023auh}
    \begin{equation}
        \frac{\D M_{\text{acc}}}{\D t} =F_{\text{cap}}\ \rho_\chi \pi R_{\text{star}}^2 \langle v_{\text{gf}}\rangle\sqrt{\frac{8}{3\pi}}\left(1+\frac{3v_e^2}{2\langle v_{\text{gf}}\rangle^2}\right).
        \label{eq:accretion rate}
    \end{equation}
$\langle v_{\text{gf}}\rangle \equiv \int\D v v f_{\text{gf}}(v)$ is the average velocity over the Maxwellian velocity distribution $f_{\rm gf}(v)$ of DM in the galactic frame. $v_e = \sqrt{{2 G M_{\text{star}}}/{R_{\text{star}}}}$ is the escape velocity of the star with $M_{\text{star}}$ and $R_{\text{star}}$ denoting the star's mass and radius respectively and $G$ is the gravitational constant. $\rho_{\chi}$ is the dark matter energy density near the star.
For the DM parameter space Eq.~\eqref{eq:DM_parameter_space} and the MS star mass range $[3,5]M_\odot$ that we are interested in, the capture probability $F_{\rm cap}$ is very close to 1, as shown in Fig.(\ref{fig:fcap}). More details are provided in the Appendix. 

If a star is located in a binary system, the DM accumulation can be enhanced by the gravitational slingshot effect from the companion star. However, this enhancement becomes significant only when the stars are in close proximity within the binary. For example, a binary of two $1.3 M_{\odot}$ stars with a short orbital period of approximately 32 hours may increase the capture rate by a factor of 1.5~\cite{Brayeur:2011yw}.  Consequently, we omit this effect from our study, as the DM capture process typically occurs long before the binary enters the close inspiral stage.

DM particles upon capture will continue to interact with the stellar matter, leading to further energy loss. Ultimately, they will thermalize with the stellar environment.
For a $\sim4M_\odot$ MS star, the thermalization timescale is related to the DM mass $m_{\chi}$ and the DM-Hydrogen scattering cross-section $\sigma_{\chi \mathrm{H}}$ as~\cite{Acevedo:2020gro} 
\begin{equation}
        t_{\text{th}} \approx 2\times 10^{-8} \mathrm{year}\left(\frac{m_\chi}{10^6 \gev}\right)\left(\frac{10^{-26}\mathrm{cm}^2}{\sigma_{\chi \mathrm{H}}}\right).
        \label{eq:thermalizationrate}
\end{equation}

For $m_{\chi}$ and $\sigma_{\chi \mathrm{H}}$ considered in the present work, the thermalization happens within a timescale much shorter than the star's lifetime. 
The thermalized DM particles become concentrated in the stellar core with a characteristic radius $r_{\text{th}}$, within which they
can form thermal bound states under the star's gravitational potential $\langle V(r)\rangle = \frac{2}{3}\pi \rho_{\text{star}} G m_\chi r^2$.\footnote{The gravitational potential of accreted DM may be neglected, as the DM energy density can only reach $\rho_\chi\sim \pi^2/12 \rho_{\text{star}}$ before the gravitational collapse of DM starts.} 
The virial theorem relates this potential to the DM thermalized kinetic energy $\langle E_k \rangle = \frac{3}{2} T_{\text{star}}$ ($T_{\text{star}}$ is the star's temperature), which implies 
$r_{\text{th}} \approx \sqrt{9 T_{\text{star}}/(4\pi G \rho_{\text{star}}m_\chi)}.$

DM within the sphere continues to accumulate until it reaches the instability threshold, triggering a collapse. For collapsing into a BH, three criteria must collectively be satisfied: Jeans instability, self-gravitating instability, and Chandrasekhar limit~\cite{Acevedo:2020gro, Ray:2023auh}. 
Detailed numerical analysis of these criteria is provided in the Appendix. For the parameter space Eq.~\eqref{eq:DM_parameter_space} and the [3,5]$M_{\odot}$ MS stars , it turns out that the Jeans Instability sets the most stringent threshold for the total DM mass $M_{\text{crit}}$ within the sphere to collapse into a BH,
\begin{equation}
\begin{aligned}\label{eq:Mcrit}
    M_{\text{crit}} =&4\times 10^{-10}M_\odot\left(\frac{T_{\text{star}}}{10^7 \mathrm{K}}\right)^{\frac{3}{2}} \left(\frac{m_\chi}{10^6 \mathrm{GeV}}\right)^{-\frac{3}{2}}
    \left(\frac{\rho_{\text{star}}}{10^{-2} \mathrm{kg}/\mathrm{cm}^3}\right)^{-\frac{1}{2}}\\
    =&1.4\times 10^{-9}M_\odot\left(\frac{M_{\text{star}}}{4M_\odot}\right)^{1.4125}\left(\frac{m_\chi}{10^6 \mathrm{GeV}}\right)^{-\frac{3}{2}}.
\end{aligned}
\end{equation}
The second equation is obtained through scaling relations between MS stars and the Sun. Specifically, the radius and the temperature are given by $R_{\text{star}} = (M_{\rm star}/M_\odot)^{0.8} R_\odot$ and $T_{\text{star}}= (M_{\rm star}/M_\odot)^{0.475} T_\odot$ in accordance with the Hertzsprung-Russell(HR) diagram. The star's lifetime scales as $\tau_{\text{star}}= (M_{\rm star}/M_\odot)^{-2.5} \tau_\odot$~\cite{lecchini2007dwarfs, hansen2012stellar}. Additionally, we assume the star's core density follows the scaling of average density, thus the core density can be written as $\rho_{\text{star}}\propto M_{\text{star}}/R_{\text{star}}^3 \propto (M_{\rm star}/M_\odot)^{-1.4} \rho_\odot$.  

To form a BH within a star's lifetime $\tau_{\text{star}}$, the accumulation time $t_{\rm acc} = M_{\rm crit}/(\mathrm{d}M_{\rm acc}/\mathrm{d}t)$ for the mini BH should satisfy $t_{\rm acc} \lesssim \tau_{\text{star}}$, with the DM accumulation rate in Eq.~\eqref{eq:accretion rate}. This sets the criterion for the DM density in the vicinity of a star 
\begin{equation}
\begin{aligned}
\rho_{\text{crit}} 
&\approx 4.10\ \mathrm{GeV/cm}^3 \left(\frac{10^6 \mathrm{GeV}}{m_\chi}\right)^{3/2}\left(\frac{M_{\text{star}}}{4M_\odot}\right)^{0.9}\left(\frac{\langle v_\text{gf}\rangle}{440\mathrm{km}/\mathrm{s}}\right).
\end{aligned}
\end{equation}

Here we assumed $3v_e^2/(2\langle v_{\text{gf}}\rangle^2)\gg 1$, a valid approximation for typical $\langle v_\text{gf} \rangle$. 

After the mini BH forms, its mass grows by accreting both stellar material and newly captured DM particles. The former is characterized by Bondi accretion \cite{Bondi:1944jm} and the latter depends on the DM capture rate.  Additionally, the mini BH may evaporate via Hawking radiation. Incorporating these processes, the mini BH mass at the stellar core evolves as
\begin{equation}\label{eq:BHeva}
\frac{\mathrm{d} M_{\mathrm{BH}}}{\mathrm{d} t}=\frac{4 \pi \rho_{\text{star}}\left(G M_{\mathrm{BH}}\right)^2}{c_{\text{star}}^3}+\frac{\D M_{\text{acc}}}{\D t}-\frac{f\left(M_{\mathrm{BH}}\right)}{\left(G M_{\mathrm{BH}}\right)^2}.
\end{equation}
$c_{\text{star}}\approx\sqrt{T_{\text{star}}/m_\mathrm{H}}$ is the speed of sound in the stellar matter. $f\left(M_{\mathrm{BH}}\right)$ is the Page factor, characterizing the strength of Hawking radiation. We take it to be $1/(74\pi)$ as the most aggressive choice, assuming the emission of all species of SM particles with gray-body corrections~\cite{Arbey:2021mbl,Acevedo:2020gro}. Hawking radiation is less important compared with the first two accretion terms in Eq.~(\ref{eq:BHeva}). To see this, we take the largest DM mass in our parameter space as shown in Eq.~\eqref{eq:DM_parameter_space}, $m_\chi = 10^9 \text{GeV}$, leading to the lightest mini BH (c.f. Eq.~(\ref{eq:Mcrit})) which implies the highest Hawking radiation rate and the lowest accretion rate. 
With this conservative choice, for a progenitor star with 4$M_\odot$, the first two terms in Eq.~(\ref{eq:BHeva}) are $4.5\times10^{28}\gev/\mathrm{s}$ and $4.2\times10^{31}\gev/\mathrm{s}$ respectively
\footnote{We note that as the accretion progresses, the BH grows larger, and the stellar matter accretion becomes dominant over the DM accretion. Additionally, the effect of Hawking radiation becomes increasingly negligible.} 
while the Hawking radiation rate is only $6.2\times10^{10}\gev/\mathrm{s}$. Taking a smaller DM particle mass, the Hawking radiation is even less important. 
Thus, for the parameter space in Eq.~\eqref{eq:DM_parameter_space}, the mini BH always persists, potentially altering the fate of the host MS star. \\

\section{Low-mass gap black hole and a possible dark matter solution}

The effect from DM capture on the evolution of stars varies depending on the stellar mass. Especially, we are interested in the stars within the mass range of $[0.5,8]M_\odot$, where the standard stellar evolution predicts the formation of WDs. In this section we discuss the influence of the mini BH to the host star. We find the DM induced mini BH inside the host star can transform the star into a LMBH as its final stage, without changing its most lifespan of the main sequence star evolution.

If no DM-induced BH forms in the center, a star within the mass range of $[0.5,8]M_\odot$ will evolve into a red giant after depleting the hydrogen fuel, eventually transforming into a WD, based on the traditional picture. If a mini BH forms during the MS phase of a star with a small mass, the accretion rate may not be significant enough to fully consume the star within the MS phase. 

To clarify, we follow ~\cite{Bellinger:2023wou,Caplan:2023ddo} to discuss the two main processes of accretion: Bondi accretion and Eddington accretion. Bondi accretion describes spherical accretion onto a BH, with its rate depending on the sound speed and density of the core of the star. It is effcient however according to ~\cite{Acevedo:2020gro}, Bondi accretion rate can only be applied to mini BHs with mass smaller than a few $10^{-9}M_\odot$  for a typical MS star with mass $\gtrsim 3M_\odot$.  After that, radiation produced during accretion counteracts the infalling matter, significantly reducing the accretion rate in what is known as Eddington accretion rate, which is much slower than the Bondi accretion rate given the same BH mass.  Notably, the radiation pressure exerted by photons depends solely on the mass accretion rate $\D M_{\rm BH}/\D t$ and properties of the infalling matter (primarily opacity), rather than on the mass of the mini BH itself. 

According to the conclusions of ~\cite{Bellinger:2023wou,Caplan:2023ddo}, a mini BH with mass $\gtrsim 10^{-10}M_\odot$ can consume a MS star with mass greater than $1M_\odot$ within several billion years.  SM star with smaller mass would require more time to accrete due to the higher opacity and lower core density, requiring substantially more time for the mini BH to fully consume a lighter host star—potentially exceeding cosmological timescales on the order of $\mathcal{O}(10 \, \mathrm{Gyr})$. Heavier SM stars accretes faster, but their lifetimes would decrease even faster with their masses. Thus the mini BHs change the lifetime of heavier stars by $< 10\%$.

On the other hand, a MS star will enter its red giant phase, where a helium core forms within it. As the red giant evolves to a late stage, electron degeneracy increases,  decreasing the opacity. By the final stages, opacity  would drop by an order of magnitude compared to the MS phase~\cite{Raffelt:1994ry,Iglesias:1996bh,Dearborn:2005bb}, significantly boosting the Eddington accretion rate. Considering these effects, it is reasonable to expect that the accretion is efficient enough in $[3, 5] M_\odot$ stars for them to end their lives as LMBHs of comparable mass, provided that dark matter successfully induces a mini black hole at the core. 
We note, the mini BH is much lighter than its hosting MS star during most of the star lifetime, but it converts the star into a LMBH as the final product.

MS stars with masses beyond $8 M_\odot$ possess the energy necessary to trigger supernova explosions, leading to the violent mass ejection and potentially generating a NS or an ABH. These massive stars have relatively short lifetimes. During the MS phase, it is unlikely for a mini BH to completely consume the whole star before supernova happens. Moreover, the supernova process is rapid and drastic, with the mini BH likely having minimal impact on the star's evolution during this phase. Hence, we expect that the evolution of these heavy MS stars remains largely unaffected even in the presence of a mini BH at the center.

It's worth noting that the  observation of stars with an given age can impose possible constraints on this model. Here we comment on three possible scenarios. 
\begin{enumerate}
\item The local observed stars. Stars located within \(\leq 3\,{\rm kpc}\) of us have been extensively observed, forming the well-known HR diagram derived from their magnitudes and temperatures ~\cite{robin2022self,widmark2022mapping}. Numerical simulations in ~\cite{Bellinger:2023wou} show that mini BHs inside solar-like stars can cause these stars to deviate from the HR diagram after a significant portion of the star’s mass has been accreted by the mini BH.  We estimate the growth time (\(t_{\rm grow}\)) of the mini-BH, from its formation to this critical mass, as the Bondi accretion time before transitioning to Eddington accretion.  The Bondi accretion time is inversely proportional to the initial mass of the mini BH, scaling as:
\begin{equation}
    t_{\rm grow} \propto \frac{1}{M_{\rm init}} \approx 1.25\,{\rm Gyr}\left(\frac{m_\chi}{10^6\,\mathrm{GeV}}\right)^{3/2},
\end{equation}
for solar-like stars with initial mini BH masses \(\leq 10^{-9} M_\odot\).  Given that these local stars experience similar dark matter densities as the Sun, we adopt a local dark matter density of \(\rho_\chi = 0.4 \mathrm{GeV/cm}^3\) and an average velocity of \(270 \mathrm{km/s}\) to estimate the accumulation time (\(t_{\rm acc}\)). To set constraints, we require that the combined time for mini BH formation and growth to significantly impact the HR diagram (\(t_{\rm acc} + t_{\rm grow}\)) be less than the maximum solar lifetime (\(\sim 10\,{\rm Gyr}\)). This leads to the following constraint on the dark matter particle mass:
\begin{equation}
    0.7\times 10^6 {\rm GeV}\leq m_\chi \leq 3.8\times 10^6{\rm GeV}.
\end{equation}
Within this range, a mini BH would form inside a local solar-like star and cause deviations from the HR diagram within the star’s lifetime. This constraint differs slightly from those reported in ~\cite{Acevedo:2020gro, Ray:2023auh}, as we account for the mini BH growth time. The resulting excluded parameter space is illustrated in Fig. (\ref{fig:eventrate}).

\item Old stars in regions of high dark matter density: The galactic center has a significantly higher dark matter density compared to the solar system, which could lead to faster mini BH formation times. However, current observations of stars in the galactic center are highly limited. For the Nuclear Star Cluster sits closely to the galactic center ($\leq 150 {\rm pc}$), only exceptionally bright, young (\(\leq 10 \, {\rm Myr}\)) and massive (\(\geq 10 M_\odot\)) stars, and the collective behavior of dimmer stars, possibly with lower masses can be identified ~\cite{Genzel:2010zy,paumard2014gcirs}. These massive stars, with their short lifetimes, are insufficient for a mini BH to both form and grow significantly, meanwhile the dimmer stars cannot be resolved individually with current observational capabilities ~\cite{Genzel:2010zy,Wegg:2013upa,hasselquist2020exploring}, leaving them both irrelevant to the scope of our study.  
For the outer regions like the galactic bulge, observations of individual stars exist but remain limited. As reviewed in \cite{bensby2017chemical},  current observations are constrained by factors such as extinction, reddening, and theoretical uncertainties in stellar atmospheres, including metallicity and surface gravity effects \cite{bica2016globular,fujii2006sio,matsunaga2016lack,puls2005atmospheric}. Meanwhile, key stellar properties to our mechanism, such as mass and age, are typically inferred statistically from limited photometric data, reducing the robustness of constraints on dark matter mechanisms \cite{zoccali2003age,longmore2013variations,gennaro2015initial}. To explore the potential of future advancements in observational tools , stars in this region to constrain non-annihilating dark matter, we also derive a projective constraint assuming a full understanding of stars within the bulge at $\sim 1\,{\rm kpc}$ from the galactic center. Using the NFW profile, we estimate a corresponding dark matter density of $\sim 6.6\,{\rm GeV}/{\rm cm}^3$ and a rotation velocity of $\sim 115\,{\rm km/s}$ based on the Milky Way’s rotation curve \cite{mroz2019rotation}. These projections inform the constraints illustrated as dark blue shaded regions in Fig. 3, surpassing existing constraints derived from objects nearby ~\cite{Acevedo:2020gro,Ray:2023auh}.

\item White dwarfs in regions of high dark matter density: Since our mechanism requires stars within our target mass range to end their life cycles as black holes, observing a white dwarf in a high dark matter density environment would impose a much stronger constraint than the measurement of the sun. However similarly as before, white dwarfs are exceptionally faint objects, and no white dwarfs have yet been observed near the galactic center due to the limitations of current observational capabilities.
\end{enumerate}  

\section{Low-mass gap Black hole distribution}

The spatial distribution of LMBHs depends on  the relationship between the DM density profile and the spatial distribution of MS stars. Galaxies of varying mass and morphology will thus exhibit different proportions of main sequence stars that transform into LMBHs. Consequently, this leads to a unique prediction for the probability distribution of these LMBHs across different galaxy masses. In this section, we estimate this distribution based on the observed analytic fit of the local ($z=0$) galaxy distribution. 

\begin{figure}
    \centering   
    \includegraphics[width=0.8\textwidth]{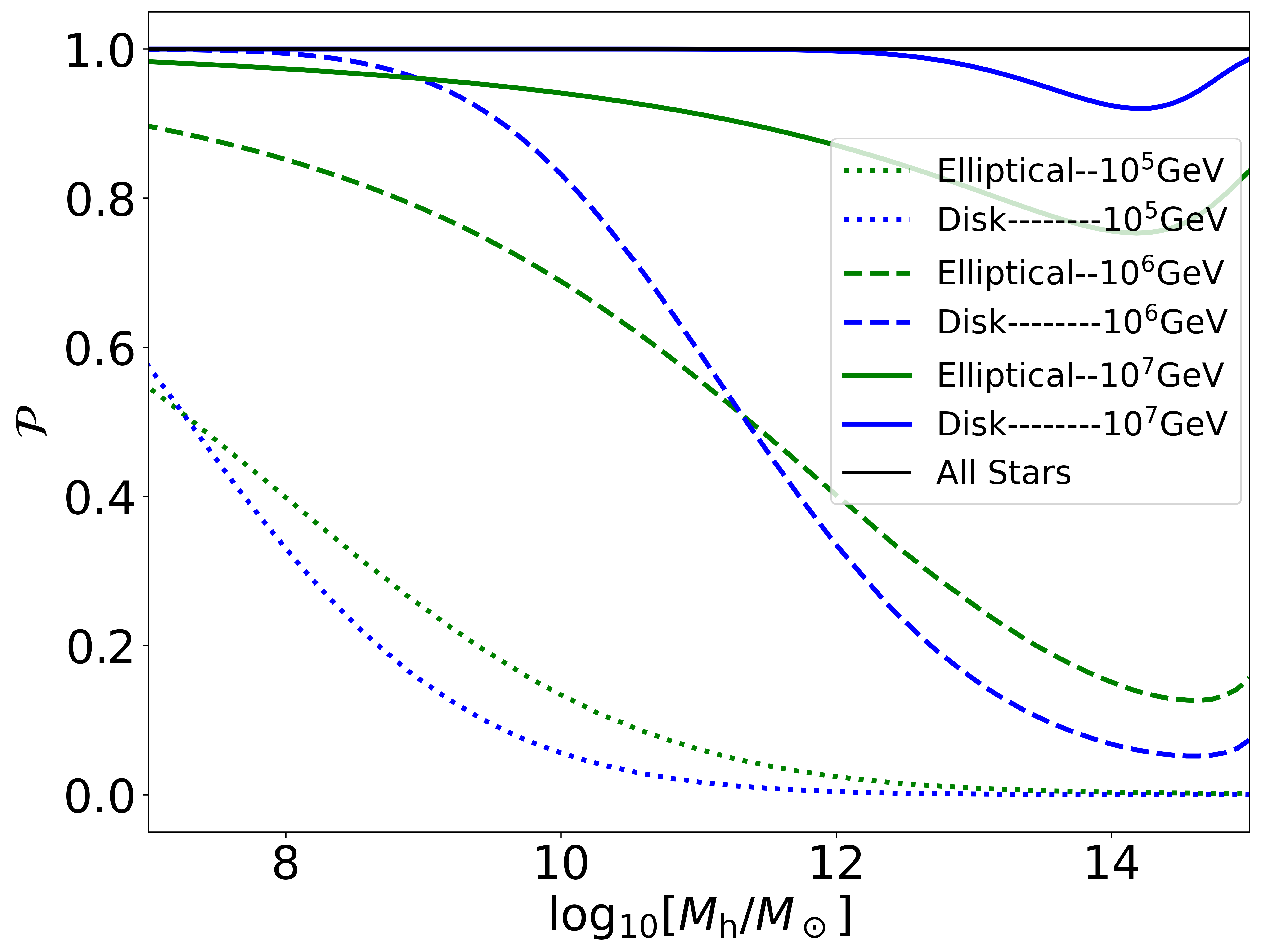}
    \caption{With various choices of DM masses and galaxy types, we show the probability for progenitor stars within mass range $[3,5]M_\odot$ to become LMBHs, as a function of the halo mass. The black line labeled as ``All Star" denotes the scenario where all MS stars within the same mass range are converted into LMBHs ($\mathcal{P}=1$). }
    \label{fig:proportionbyxplot}
\end{figure}

For the spatial distribution of the DM density $\rho_\chi(\vec{r}_g)$, we take the NFW profile~\cite{Navarro:1996gj}, normalized to the galaxy halo mass $M_{\rm h}$. The total stellar mass is related to $M_{\rm h}$ through a relationship derived from the Bolshoi-Planck simulation \cite{Behroozi:2019kql,Cui:2021hlu}. Additionally, the spatial distribution of stars $\rho_s(\vec{r}_g)$ depends on the type of the galaxy, either disk or elliptical. The detailed descriptions of these quantities are provided in the Appendix. 
With everything prepared, we can calculate the probability of a star with mass between $[3,5]M_\odot$ within a galaxy of mass $M_{\text{h}}$ becoming a LMBH during its lifetime, 
\begin{equation}\label{eq:prob-Mhalo}
\mathcal{P}(M_{\text{h}}) = \frac{N_{\text{BH}}^{\text{LM}}(M_{\text{h}})}{N_{\text{star}}^{\text{LM}}(M_{\text{h}})}.
\end{equation}
The denominator is the total number of stars with mass between $[3,5]M_\odot$,
\begin{equation}
    N_{\text{star}}^{\text{LM}}(M_{\text{h}})=\int_{3M_\odot}^{5M_\odot} \frac{\D M_{\text{star}}}{\langle M_{\text{star}}\rangle}\frac{\D n_{\text{star}}}{\D M_{\text{star}}}\int\D V \rho_s(\vec{r}_g).
\end{equation}
Here the volume integration extends to the virial radius of the galaxy $R_{\text{vir}}(M_{\text{h}})$. We take the star mass distribution $\frac{\D n_{\text{star}}}{\D M_{\text{star}}}$ following the initial mass function (IMF) ~\cite{Kroupa:2000iv,Cui:2021hlu},
\begin{equation}\label{eq:IMF}
        \frac{\D n_{\text{star}}}{\D M_{\text{star}}}\propto M_{\text{star}}^{-2.3}, \ \ 0.5 M_{\odot} \leq M_{\text{star}}<100 M_{\odot}.
    \end{equation}
The IMF is normalized by the averaged star mass $\langle M_{\text{star}} \rangle = \int \D M_{\text{star}}\frac{\D n_{\text{star}}}{\D M_{\text{star}}} M_{\text{star}}$, whose integration range runs for all star masses. This normalization factor will eventually be canceled in our calculation. Additionally, the numerator of Eq.~(\ref{eq:prob-Mhalo}) is the number of stars in the same mass range but will become LMBHs through DM capture,
\begin{equation}
\begin{aligned}
    N_{\text{BH}}^{\text{LM}}(M_{\text{h}})=&\int_{3M_\odot}^{5M_\odot} \frac{\D M_{\text{star}}}{\langle M_{\text{star}}\rangle}\frac{\D n_{\text{star}}}{\D M_{\text{star}}}\,\times\\
    &\int \D V\rho_s(\vec{r}_g)\Theta\left[\rho_\chi(\vec{r}_g)-\rho_{\text{crit}}(\vec{r}_g,M_{\text{star}})\right].
\end{aligned}
\end{equation}
In Fig.~\ref{fig:proportionbyxplot}, we take several benchmarks for DM masses and show the probability of a star with mass between [3,5] $M_\odot$ to become a LMBH as a function of the halo mass. We see that almost all stars in this mass range could become LMBHs if the DM mass is large enough. This is also consistent with the DM mass upper limit imposed by the survival of the Sun.

\begin{figure}
    \centering   \includegraphics[width=0.8\textwidth]{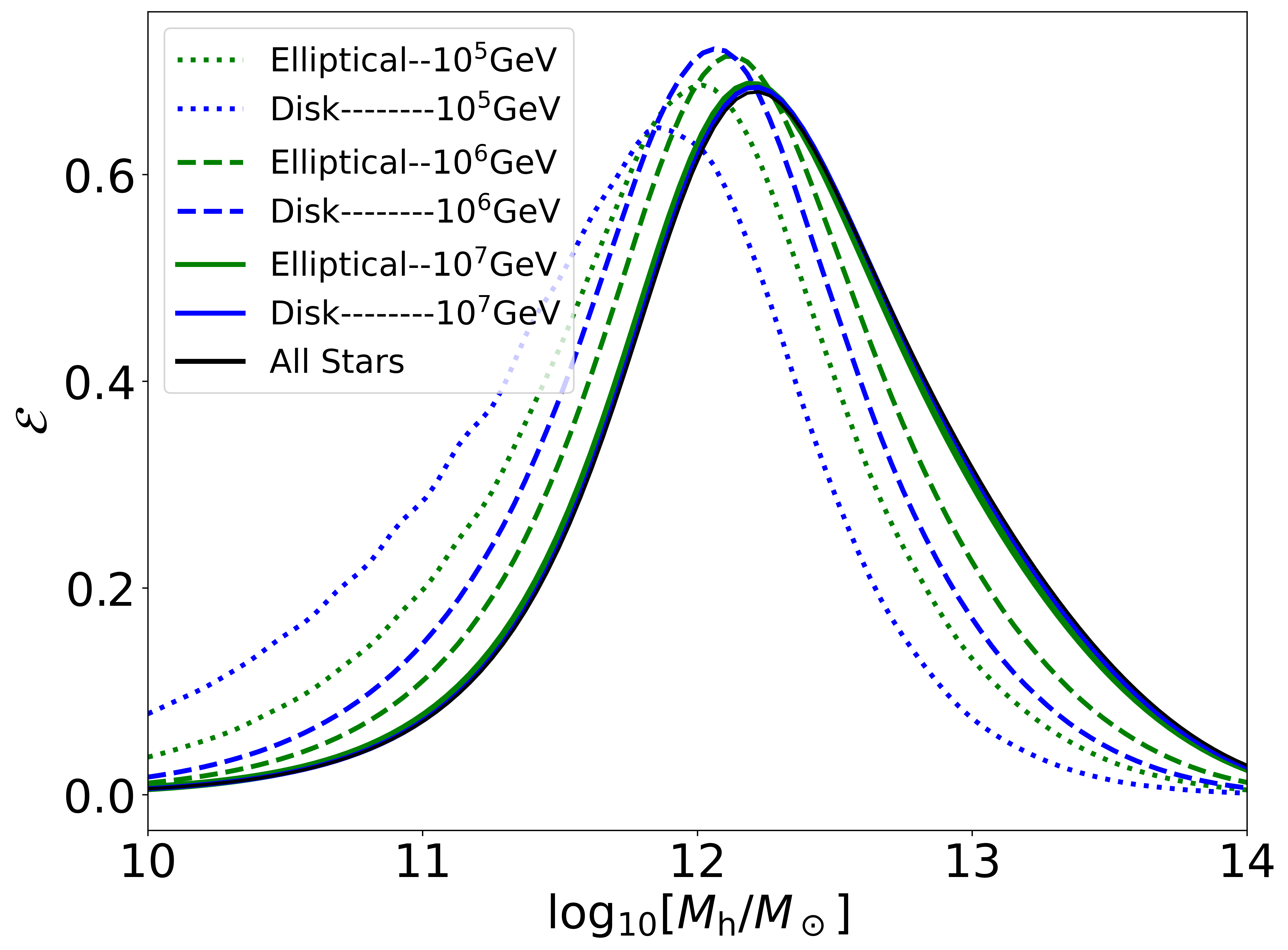}
    \caption{With various choices of DM masses and the galaxy types, we show the relative probability distribution for $[3,5]M_\odot$ LMBHs as a function of the logarithmic halo mass. The black line labeled as ``All Star" denotes the scenario where all MS stars within the same mass range are converted into LMBHs.}
    \label{fig:WeightedEbyxplot}
\end{figure}

Next, we calculate the relative probability distribution of LMBHs as a function of logarithmic halo mass,
\begin{equation}\label{eq:eps}
\mathcal{E}(M_{\text{h}}) =  \frac{\frac{\D n}{\D \log{[M_{\text{h}}]}}N_{\text{BH}}^{\text{LM}}(M_{\text{h}})}{\int \D \log{[M_{\text{h}}]}\frac{\D n}{\D \log{[M_{\text{h}}]}}N_{\text{BH}}^{\text{LM}}(M_{\text{h}})}.
\end{equation}
Here the halo mass distribution  $\frac{\D n}{\D \log{[M_{\text{h}}]}}$ is taken as the Sheth-Tormen distribution at $z=0$ \cite{Sheth:1999mn,Cui:2021hlu}. We show the result of $\mathcal{E}(M_{\text{h}})$ in Fig.~\ref{fig:WeightedEbyxplot}.
We emphasize that the relative probability distribution, $\mathcal{E}(M_{\text{h}})$ in Eq.(\ref{eq:eps}), is a unique prediction in this mechanism, which serves as an excellent discriminator to distinguish this mechanism from other LMBH formation mechanisms.

\section{Binary event rate estimation}

One useful way to search for a LMBH in a distant galaxy is through GW radiation emitted when it merges with a compact object. The waveform during the merger offers valuable insights for identifying the properties of the binary system, especially the masses involved.
In this section, we take LMBH-NS binaries as a benchmark and estimate their merger rate.

The formation and evolution of binaries involve several astrophysical uncertainties, leading to large error bars in the predictions for binary merger rates. For instance, the merger rate for an ABH-NS binary can vary widely, from $0.1$ to $800 \mathrm{Gpc}^{-3}\mathrm{yr}^{-1}$ for isolated binaries~\cite{Belczynski:2001uc,Samsing:2020qqd,Broekgaarden:2021iew,LIGOScientific:2021qlt}, and from $0.1$ to $100 \mathrm{Gpc}^{-3}\mathrm{yr}^{-1}$ for binaries in dynamical environments such as globular clusters~\cite{PortegiesZwart:1999nm,Antonini:2013tea,Hoang:2020gsi,Sedda:2020wzl,LIGOScientific:2021qlt}.

In order to reduce the uncertainty, we compare the merger rates of two distinct binary systems: the LMBH-NS binary and the ABH-NS binary. The masses of LMBH and ABH are related to the corresponding progenitor stars in different ways. The LMBH originates from a progenitor star with a mass between $[3,5]M_\odot$ induced by DM. The LMBH and its progenitor star are approximately equal in mass, which is motivated by the expectation that no drastic explosion should occur when such a low-mass MS star evolves to its final stage
\footnote{Particularly, when the DM mass $m_\chi<10^6\gev$, the mini BH can swallow the progenitor star within 1 Gyr~\cite{Bellinger:2023wou,Caplan:2023ddo}, before it turns into a red giant.}.
While the ABH arises from a progenitor star with a mass between $[20,80]M_\odot$ which loses a significant fraction of mass through supernova explosion, resulting in an ABH heavier than $5M_\odot$ and lighter than approximately $20M_\odot$. As demonstrated in Refs.~\cite{Vink:2001cg,Vink:2011kd,10.1093/mnras/stv1281,Cui:2021hlu}, the ABH mass correlates with the original MS star mass by a factor of O(1), typically centered around 4. To simplify our estimation, we take $M_{\text{ABH}} = M_{\text{star}}/4$.

To compare the LMBH-NS rate with the ABH-NS rate, several assumptions need to be made.
The mass distribution of progenitor stars follows the IMF described in Eq.~(\ref{eq:IMF}). We assume that this scaling is preserved in binary systems containing a NS before the progenitor star evolves into a BH.
Moreover, during the ABH formation, the supernova explosion typically leads to a significant kick. Such effects could mildly increase the merger rate, as the kick induces a non-trivial eccentricity, accelerating the energy loss rate via gravitational waves. However, the kick may also create too much kinetic energy, causing the binary to become unbound, thereby decreasing the merger rate~\cite{Belczynski:2001uc,Tang:2019qhn,LIGOScientific:2021qlt}. Overall, this effect may alter the merger rate of ABH-NS binaries by a factor ranging from approximately 0.1 to 1.2. For an order-of-magnitude estimation, we consider this factor to be 1. Under these assumptions, we find that the number density of LMBH-NS binaries is approximately 6.84 times greater than that of ABH-NS binaries.

At last, a BH-NS binary with a fixed initial orbit separation merges faster for a heavier BH due to a higher GW emission rate. This affects the measured merger rates for a GW detector with an O(1)~yr observation period. 
The maximum orbital period for a BH-NS binary to merge within $\Delta t^{\rm ob} = 1$ year under the post-Newtonian approximation is~\cite{Maggiore:2007ulw}
\begin{equation}
    P^{\rm ob}_{\rm max}(M_{\rm BH}) \approx\left(\frac{256\pi^{8/3}\Delta t^{\rm ob}G^{5/3}M_{\rm BH}M_{\rm NS}}{5 (M_{\rm BH}+M_{\rm NS})^{1/3}}\right)^{8/3}. 
\end{equation} 
In~\cite{tokovinin2020formation}, the probability distribution of close binary periods $\frac{\mathrm{d} n}{\mathrm{d}\log P}$ follows a Gaussian distribution, centered at $\log_{10}(P/\text{day}) = 4.8$, with a dispersion of approximately 2.3. This distribution is universal across all BH masses. Taking $M_{\rm NS}=1.5M_\odot$, the differences in binary number densities and merger efficiencies together yield a fudge factor $A\simeq 5.79$ when we relate the observed merger rates of ABH-NS binaries and LMBH-NS binaries. More details can be found in the Appendix.\ref{sec:Relative-Binary-Rate}.

\begin{figure}
    \centering   \includegraphics[width=0.8\textwidth]{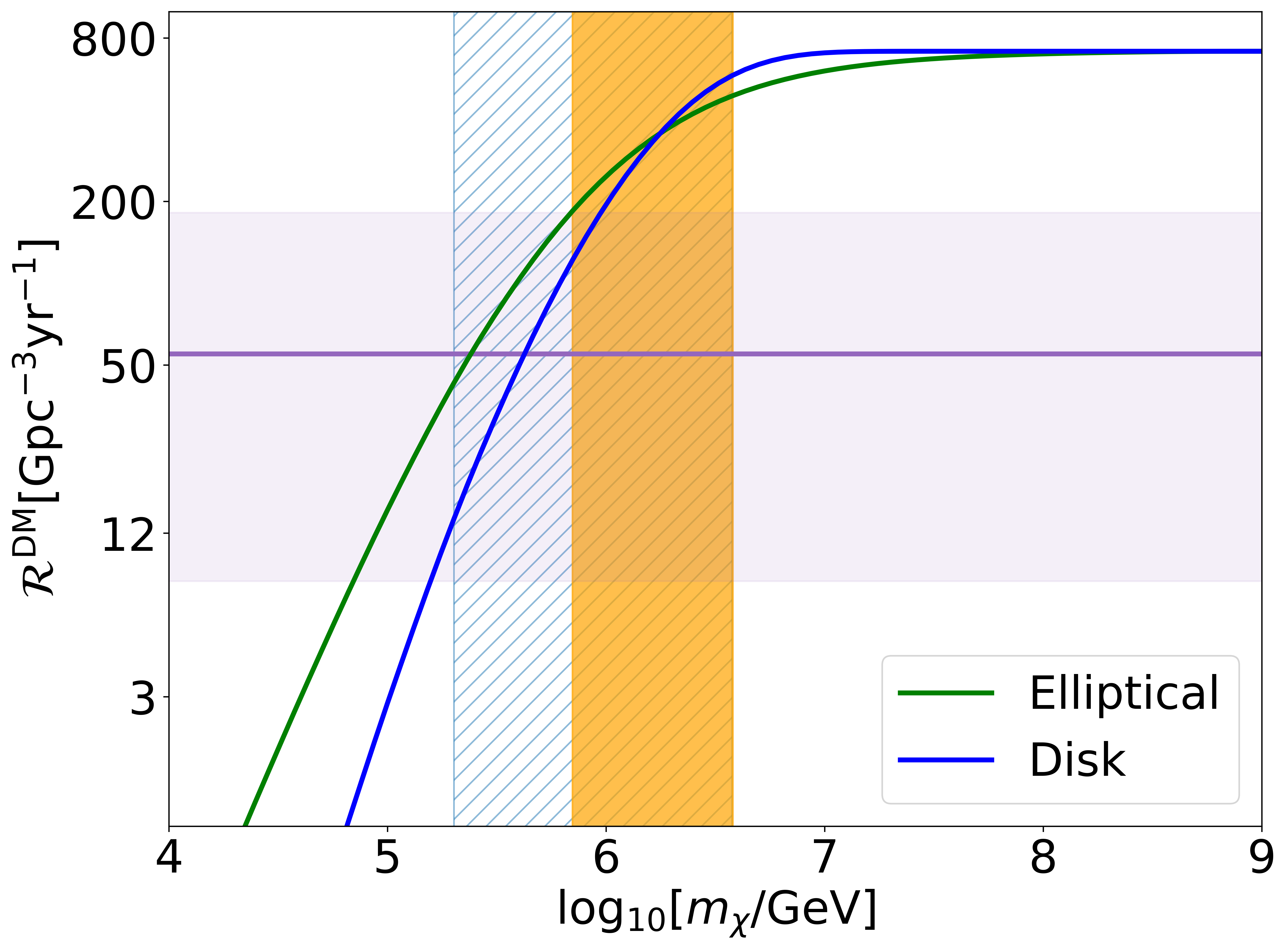}
    \caption{For different galaxy types, we show the predicted LMBH-NS merger rate as a function of DM mass. In comparison, the purple line (band) corresponds to the merge rate (uncerntainty) by LVK's recent observations~\cite{LIGOScientific:2024elc}. The orange region is excluded, which is the constraint from the observed local solar-like stars with maximum age ($10\,{\rm Gyr} $) not deviated from the HR diagram. The dark blue shaded regions indicates projected constraints if one observes a solar-like star with its maximum age at $1\,{\rm kpc}$ from the galactic center.}

    \label{fig:eventrate}
\end{figure}

Consequently, after the convolution with the halo mass distribution, one can write the rate ratio of these two types of mergers as
\begin{equation}
    \frac{\mathcal{R}^{\text{LMBH}}}{\mathcal{R}^{\text{ABH}}} = A\frac{ \int \D \log{[M_{\text{h}}]}\frac{\D n}{ \D \log{[M_{\text{h}}]}}N_{\text{BH}}^{\text{LM}}(M_{\text{h}})}{\int  \D \log{[M_{\text{h}}]}\frac{\D n}{\D \log{[M_{\text{h}}]}}N_{\text{star}}^{\text{HM}}(M_{\text{h}})}.
\end{equation}
Here $N_{\text{star}}^{\text{HM}}=\int_{20M_\odot}^{80M_\odot} \frac{\D M_{\text{star}}}{\langle M_{\rm star}\rangle}\frac{\D n_{\text{star}}}{\D M_{\text{star}}}\int\D V \rho_s(\vec{r}_g)$, which is the total number of heavy stars given a galaxy halo mass.
Taking the central value of the ABH-NS merger rate measured by LVK as $\mathcal{R}^{\text{ABH}}=130\mathrm{Gpc}^{-3}\mathrm{yr}^{-1}$~\cite{LIGOScientific:2021qlt}, we show our predicted LMBH-NS binary merger rate as a function of DM mass in Fig.~\ref{fig:eventrate}. It is worth mentioning that a DM with mass smaller than $10^6$ GeV gives a LMBH-NS merger rate consistent with the recent LVK's reported rate~\cite{LIGOScientific:2024elc}, $55_{-47}^{+127} \mathrm{Gpc}^{-3} \mathrm{yr}^{-1}$, shown as the purple band in Fig.~\ref{fig:eventrate}.

\section{Conclusion and Discussions}
The detection of the LMBH-NS merger by LVK indicates the existence of LMBHs. This is a surprising result and may require a detailed study on their possible formation mechanism. 
We find that the strongly interacting DM model offers a potential explanation for the formation of these LMBHs. A significant portion of the parameter space gives a LMBH-NS merger rate consistent with LVK's measurement.

Moreover, whether a mini BH can form within a star and eventually convert it into a LMBH depends on the properties of the DM in the star's vicinity. This leads to a unique feature in the LMBH distribution as discussed in Eq. (\ref{eq:eps}). Currently, the GW network is still in its beginning stage, and the angular resolution is not good enough to identify the host galaxy of the merger on an event-by-event basis. However, with the expansion of the GW detector network and enhancements in individual detector sensitivity, we expect more LMBH-NS mergers can be measured. Both event statistics and angular resolution have the potential for significant improvements. If the host galaxy can be identified, the distribution of LMBH-NS mergers can be directly tested. Even without event-by-event identification of the host galaxy, statistical analyses can be conducted based on the distribution of potential host galaxies from surveys like the Dark Energy Spectroscopic Instrument~\cite{DESI:2022xcl,DESI:2023dwi}. It is conceivable that such LMBH-NS distribution could be utilized to verify or refute the LMBH formation mechanism studied here.

\appendix

\begin{center}
\textbf{\large Appendices: Dark Matter-Induced Low-Mass Gap Black Hole Echoing LVK Observations}
\end{center}

\setcounter{equation}{0}
\setcounter{figure}{0}
\setcounter{table}{0}
\setcounter{section}{0}
\makeatletter
\renewcommand{\theequation}{A\arabic{equation}}
\renewcommand{\thefigure}{A\arabic{figure}}
\renewcommand{\bibnumfmt}[1]{[#1]}
\renewcommand{\citenumfont}[1]{#1}

\hspace{5mm}

\renewcommand{\thefigure}{A.\arabic{figure}} 
\setcounter{figure}{0}

This Appendix details the mechanisms of dark matter capture within stellar bodies. It also provides explicit formulas for the criteria of gravitational collapse, as mentioned in the main text, elucidating the conditions under which such collapses are expected to occur. Additionally, the material offers insights into the galaxy number density distribution with respect to halo mass, as well as the distribution of stellar and dark matter within each galaxy. 
Finally, the modifications to the relative binary formation rate between two kinds of binaries (1. binaries of low-mass gap black hole and neutron star, and 2. binaries of astrophysical black hole and neutron star) are meticulously calculated.\\

\section{Dark matter capture rate in the star body}

\par In this section, we follow the calculations presented in ~\cite{Gould:1987ir,Jungman:1995df,Acevedo:2020gro,Ray:2023auh}. 
We simplify the interactions between DM and stellar matter as one-dimensional head-on collisions. This simplification provides a reasonable estimation of the orders of magnitude, as demonstrated in ~\cite{Bramante:2018qbc,Bramante:2019fhi,Acevedo:2020gro}.
We first estimate the typical energy loss for each collision between a DM particle, denoted as $\chi$, and hydrogen, the predominant component within an MS star. Assuming the initial velocity of the dark matter particle as $v_{\text{i}}$, the final velocity of the DM after a single scattering event with a hydrogen can be expressed as $v_{\text{f}} = v_{\text{i}}\sqrt{1-z\beta}$, where $\beta$ is defined as $\beta \equiv 4m_{\rm H} m_{\chi}/(m_{\rm H} + m_{\chi})^2$ and $z$ is a geometric factor averaged to $z=1/2$ for isotropic spin-independent collision angles.
For the parameter space that we are interested in (refer to Eq.~\eqref{eq:DM_parameter_space}), DM mass is always much larger than the mass of hydrogen, which leads to $\beta \approx 4\gev/m_\chi$.
When DM particles pass through the star’s interior, the average number of collisions can be estimated as
    \begin{equation}
        N(\theta) = \int_0^L n_{\rm H}(r)\sigma_{\chi \rm H}\D L.
    \end{equation}
Here, $\theta$ represents the angle between the velocity of the DM particle and the radial direction $\hat{r}$ of the star upon entry. Neglecting any change in the DM propagation direction, the distance traveled by the DM particle within the star can be expressed as $L=2R_{\text{star}}|\cos{\theta}|$, where $R_{\text{star}}$ denotes the radius of the star.

Additionally, $n_{\rm H}(r)$ denotes the hydrogen number density at a distance $r$ from the center of the star, while $\sigma_{\chi \rm H}$ represents the cross-section for DM-hydrogen interaction. Therefore, with a fixed angle $\theta$, the maximum initial velocity for a DM particle to be captured is 
    \begin{equation}\label{eq:v_max}
        v_{\max}(\theta) = \frac{v_e}{\left(1-z \beta_{\rm H}\right)^{N(\theta) / 2}}.
    \end{equation}
Here $v_e$ is the escape velocity of the star, $v_e = \sqrt{{2 G M_{\text{star}}}/{R_{\text{star}}}}$, with $M_{\text{star}}$ and $R_{\text{star}}$ represent the mass and radius of the star, respectively.

For the DM velocity distribution, we adopt a Maxwellian distribution in the galactic frame, given by
\begin{equation}
        f_{\text{gf}}(v) = \frac{4}{\sqrt{\pi}}\frac{v^2}{v_0^3}\mathrm{exp}{\left(-\frac{v^2}{v_0^2}\right)},
        \label{eq:MBdistribution}
    \end{equation}
    where $v_0$ is taken to be the circular velocity $v_{\text{cir}}$ in the halo, which should also be the virial velocity $v_{\text{vir}}$ according to Virial Theorem.
    When considering a star moving at velocity $\vec{v}_{\rm star}$ relative to the galactic center, the DM velocity in the star's frame requires adjustment through a Galilean transformation.    
    Moreover, as DM particles approach the surface of the star, they experience acceleration due to the gravitational potential. For a DM particle with velocity $\vec{u}$ at infinity in the star's frame, its velocity upon entering the stellar region becomes $v^2 = |\vec{u}|^2 + v_e^2$. 

We consider a star velocity $|\vec{v}_{\rm star}| = v_0$ within the galactic frame. Consequently, the velocity distribution of DM particles at infinitely far in the star frame, $f(u)$, can be derived through a Galilean transformation of Eq.~(\ref{eq:MBdistribution}), which gives
    \begin{equation}
        f(u,\phi) = \frac{1}{N^*}\frac{u v_{\text{gf}}^2(u)}{v_{\text{gf}}(u)-v_0\cos\phi}e^{-v_{\text{gf}}^2(u)/v_0^2},
    \end{equation}
    where $\phi$ is the isotropic angle between the dark matter velocity $\vec{v}_{\text{gf}}$ in the galactic frame and the star's velocity ${\vec{v}_{\rm star}}$, in the galactic frame. The term $v_{\text{gf}}$ satisfies the Galilean transformation:
    \begin{equation}
       v_{\text{gf}}^2+v_0^2-2v_{\text{gf}}v_0\cos\phi=u^2.
    \end{equation}
    Here, $N^*$ is the normalization factor ensuring $\int\D u\int\D\cos{\phi} f(u,\phi)=1$.
    
\par
\begin{figure}
    \centering
    \includegraphics[width=0.5\linewidth]{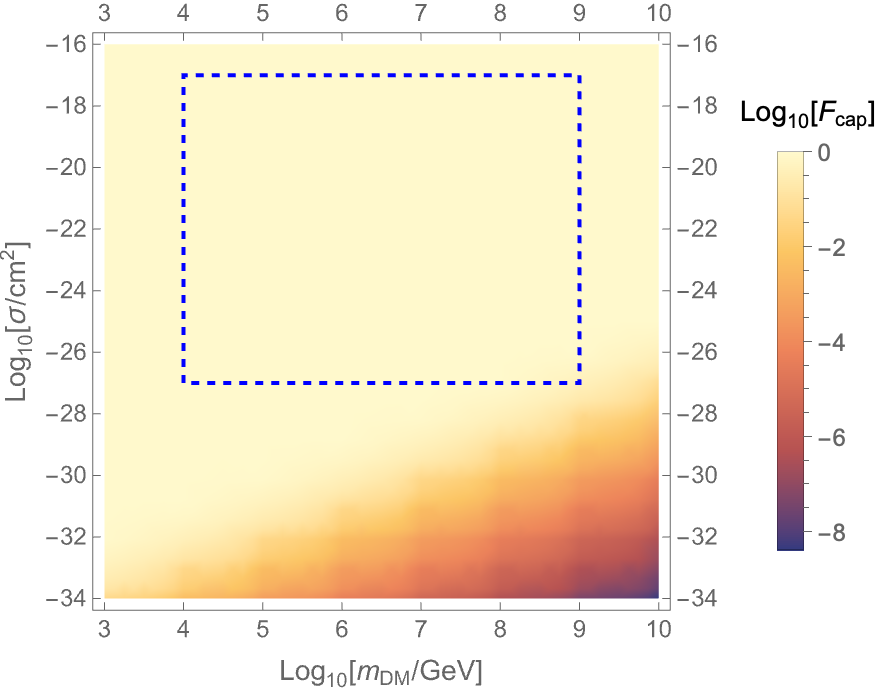}
    \caption{We show the capture probability $F_{\rm cap}$ for a $4 M_\odot$ main sequence star with different dark matter parameters. The star's hydrogen density is taken as an uniform distribution. The blue dashed square shows the parameter region of interest in Eq. (\ref{eq:DM_parameter_space}).}
    \label{fig:fcap}
\end{figure}

 As described by Eq.~\eqref{eq:v_max}, DM particles slower than the maximum initial velocity $v_{\max}(\theta)$ will be captured, eventually falling into the star's core. Thus, the maximum velocity at an infinite distance in the star frame can be derived from a gravitational acceleration:
\begin{equation}
    u_{\max}(\theta) = \sqrt{v_{\max}(\theta)^2-v_{\text{esc}}^2}
\end{equation}
Then, we derive the average capture rate for dark matter particles, $F_{\text{cap}}$, which is the fraction of DM particles captured by the star:
\begin{equation}
    F_{\text{cap}} = \int_{0}^{1}\D\cos\theta \int_{0}^{ u_{\max}(\theta)}\D u \int_{-1}^{1}\D\cos\phi f(u,\phi).
    \label{eq:CaptureFraction}
\end{equation}
We consider DM in the relatively strong interacting region, so that $F_{\rm cap}$ is very close to 1 for the parameter space of interest (Eq. \ref{eq:DM_parameter_space}). We show an example of $F_{\rm cap}$ in Fig. \ref{fig:fcap}, where the star is taken to be a $4M_\odot$ star with its mass and radius following the main sequence scaling relation provided in the main text. The dark matter distribution is taken to be the local Boltzmann Distribution with a virial velocity $230{\rm km}/{\rm s}$. The star's density profile is taken to be uniformed for simplicity. The detailed calculations considering the star's density distribution do not make a significantly difference in conclusion, and one can find them in ~\cite{Acevedo:2020gro}\\

\section{Criterion of gravitational collapse}
\par As outlined in the main text, the dark matter accreted within the star can collapse into a black hole only if it satisfies the following three criteria: Jeans instability, self-gravitating instability, and the Chandrasekhar limit. We follow ~\cite{Acevedo:2020gro, Ray:2023auh} and discuss these criteria individually below.
\par In the stable thermal state, before the onset of any instability, dark matter particles reach a virialized distribution with a temperature equal to that of the host star, $T_{\text{star}}$. At this juncture, the dark matter density is sufficiently low to be considered negligible, allowing the star's gravitational potential to be approximated as $V(r)= \frac{2}{3}\pi \rho_{\text{star}} G m_\chi r^2$. According to the Virial Theorem, this potential $\langle V(r)\rangle$ correlates with the DM kinetic energy $\langle E_k \rangle = \frac{3}{2} T_{\text{star}}$, providing an estimate for the thermal radius $r_{\text{th}}$, within which dark matter can remain gravitationally bound in a virialized thermal state:
    \begin{equation}
        r_{\text{th}} \approx \sqrt{\frac{9 T_{\text{star}}}{4\pi G \rho_{\text{star}}m_\chi}}.
    \end{equation}
\paragraph{Jeans Instability} Jeans instability occurs when the dark matter self-gravitating free-fall time $t_{\rm ff}$—the time it takes for a dark matter particle to fall from the thermal radius to the star's center—equals the sound-crossing time $t_{\rm sc}$, assuming a sound speed $c_s = \sqrt{\frac{T_{\rm star}}{m_\chi}}$. The free-fall time, given the star's harmonic gravitational potential, is:
\begin{equation}
    t_{\rm ff} = \sqrt{\frac{3\pi}{16\rho^{\rm star}_\chi G}},
\end{equation}
where $\rho^{\rm star}_\chi$ represents the dark matter density within the thermal sphere of the star. The sound-crossing time can be expressed as:
\begin{equation}
    t_{\rm sc} = \frac{r_{\rm th}}{c_s} = \frac{3}{\sqrt{4\pi G \rho_{\rm star}}}.
\end{equation}
Equating $t_{\rm ff}$ and $t_{\rm sc}$ yields:
\begin{equation}
    \rho^{\rm star}_\chi = \frac{\pi^2}{12}\rho_{\rm star}, 
\end{equation}
leading to a critical mass for the onset of Jeans Instability:
\begin{equation}
 M^{\text{JI}}_{\text{crit}}(m_\chi) = \sqrt{\frac{9\pi^3 T_{\text{star}}^3}{64 \rho_{\text{star}}G^3m_\chi^3}}.    
\end{equation}

\paragraph{Self-gravitating Instability} When the mass of accreted dark matter, denoted as $M_{\rm acc}$, becomes significantly large, the dark matter cannot maintain a virialized state at the star's temperature. In the absence of sufficient pressure from other interactions, such as the thermal pressure accounted for in the Jeans Instability scenario or the quantum pressure in the Chandrasekhar limit, the dark matter would collapse to form a black hole. The virial theorem for a dark matter particle in a bound state considers both the gravitational potential from the stellar and dark matter contributions as:
\begin{equation}
3T_{\rm star} = \frac{4}{3}\pi r^2\rho_{\rm star}G m_\chi + \frac{G M_{\rm acc}m_\chi}{r}.
\end{equation}
A solution for $r$ becomes infeasible when $M_{\rm acc}$ reaches a critical threshold, indicating that no thermal bound state can be formed within radius $r$ without considering additional interactions, signifying gravitational instability. The $r.h.s.$ above has a minimum value at $r = \left(\frac{3 M_{\rm acc}}{8\pi \rho_{\rm star}}\right)^{1/3}$, which sets the condition for self-gravitating instability:
\begin{equation}
    3T_{\rm star} =\frac{3}{2} Gm_\chi \left(\frac{8}{3}\pi \rho_{\rm star} M_{\rm acc}^2\right)^{1/3},
\end{equation}
leading to the critical mass for instability:
\begin{equation}
     M^{\text{SG}}_{\text{crit}}(m_\chi) = \sqrt{\frac{3 T_{\text{star}}^3}{\pi \rho_{\text{star}}G^3m_\chi^3}}.
\end{equation}
\paragraph{Chandrasekhar Limit} The Chandrasekhar limit defines the mass threshold beyond which quantum degenerate pressure can no longer counteract self-gravity. This criterion must be met to prevent collapse under quantum pressure. For fermionic dark matter, the critical mass is:
\begin{equation}
    M^{\text{Ch-F}}_{\text{crit}}(m_\chi)\sim \frac{M_{\rm pl}^3}{m_\chi^2},
\end{equation}
and for bosonic dark matter:
\begin{equation}
    M^{\text{Ch-B}}_{\text{crit}}(m_\chi)\sim \frac{M_{\rm pl}^2}{m_\chi},
\end{equation}
where $M_{\rm pl}\sim G^{-1/2}$ represents the Planck mass.
\par The critical mass conditions for Jeans instability and self-gravitating instability show similar parameter dependencies, albeit with slight differences in coefficients. Therefore, to ascertain the critical mass for dark matter collapse, one must compare the conditions for Jeans instability and the Chandrasekhar limit. Considering a star of mass $3M_\odot$ as the least massive star under consideration, the critical mass for Jeans instability, $M^{\text{JI}}_{\text{crit}}(m_\chi) \approx10^{48}\gev\left(\frac{m_\chi}{10^6\gev}\right)^{-3/2}$ for fermionic dark matter, and $M^{\text{JI}}_{\text{crit}}(m_\chi) \approx1.8\times10^{45}\gev\left(\frac{m_\chi}{10^6\gev}\right)^{-2}$. The scaling relations for temperature and core density in the main sequence stars indicate that an increase in star mass raises the critical mass for Jeans instability but does not alter the Chandrasekhar limit. Additionally, bosonic dark matter exhibits a lower Chandrasekhar limit. Consequently, the Jeans instability criterion predominantly determines the critical mass for dark matter to collapse into a black hole, $M_{\text{crit}}=M^{\text{JI}}_{\text{crit}}$, across most of the parameter space of interest.\\

\section{Halo mass distribution, dark matter halo, and stellar matter distribution}

We utilize the Sheth-Tormen distribution at $z=0$ for modeling the local galaxy halo mass distribution~\cite{Sheth:1999mn,Cui:2021hlu}:
\begin{equation}
    M_{\text{h}} \frac{\D n}{\D M_{\text{h}}}=\Omega_{\mathrm{m}, 0} \rho_{\mathrm{cr}, 0} \frac{\D \sigma(M_{\text{h}})}{\sigma(M_{\text{h}}) \D M_{\text{h}}} f(\sigma),
\end{equation}
where $\Omega_{\mathrm{m}, 0}\equiv \rho_{\mathrm{m},0}/\rho_{\mathrm{cr},0}\approx 0.3$ represents the current matter energy fraction, and $\rho_{\mathrm{cr},0}\approx 10^{-26}\mathrm{kg/cm}^3$ denotes the current critical energy density. The term $\sigma(M_{\text{h}})$ indicates the current root-mean-square (rms) density fluctuation, which is modeled using numerical simulations from the Bolshoi simulations, incorporating observational parameters from WMAP5 and WMAP7 data ~\cite{Sheth:2001dp,Klypin:2010qw,Prada:2011jf}:
\begin{equation}
\begin{aligned}
\sigma(M_{\text{h}}) & =\frac{16.9 y^{0.41}}{1+1.102 y^{0.20}+6.22 y^{0.333}}, \\
y & \equiv\left[\frac{M_{\text{h}}}{10^{12} h^{-1} M_{\odot}}\right]^{-1}.
\end{aligned}
\end{equation}
The function $f(\sigma)$ represents the modified analytic fit of first-crossing distribution function, which assumed to be an adaptation of the original Press-Schechter function, refined by ~\cite{Klypin:2010qw}:
\begin{equation}
\begin{aligned}
  f(\sigma)=&A_f \sqrt{\frac{2 b_f}{\pi}}\left[1+\left(\frac{b_f}{\sigma^2}\right)^{-0.3}\right] \frac{1}{\sigma} \exp \left(-\frac{b_f }{2\sigma^2}\right), \\
&A_f=0.322, \quad b_f=2.01.
\end{aligned}
\end{equation}

\par The structure of the dark matter halo is characterized by a Navarro-Frenk-White (NFW) profile~\cite{Navarro:1996gj}:
\begin{equation}
    \rho_{\chi}(r_g) = \frac{\rho_0}{r_g / R_s\left(1+r_g / R_s\right)^2},
\end{equation}
where $r_g$ is the radial distance from the center of the halo, and $R_s$ is the scale radius linked to the virial radius $R_{\text{vir}}$ through the concentration parameter $C(M_{\rm h}) = R_{\text{vir}}/R_s$. The concentration parameter, based on the Bolshoi and MultiDark simulation data for $z\approx0$, is described as~\cite{Prada:2011jf}:
\begin{equation}
\begin{aligned}
    C(M_{\rm h}) = &A_C\left[\left(\frac{\sigma(M_{\text{h}}) }{b_C}\right)^{c_C}+1\right] \exp \left(\frac{d_C}{\sigma(M_{\text{h}})^2}\right),\\
    A_C=2.881,\quad b_C=&1.257,\quad c_C=1.022,\quad d_C=0.06.
\end{aligned}
\end{equation}
The virial radius is defined as the radius within which the average density of the halo is $\Delta$ times the critical energy density, with $\Delta$ typically set to 200~\cite{White:2000jv}:
\begin{equation}
    R_{\text{vir}} = \sqrt[3]{\frac{3 M_{\rm h}}{4\pi \Delta\rho_{\text{cr,0}}}}.
\end{equation}
Consequently, $R_s$ is expressed as a function of the halo mass $M_{\rm h}$, allowing for the determination of $\rho_0$ through the normalization of the halo mass:
\begin{equation}
    M_{\rm h} = \int_0^{R_{\text{vir}}} \rho_{\chi}(r_g) 4\pi r_g^2 \D r_g. 
\end{equation}
However, it's noted that the Milky Way's halo does not perfectly align with this model due to its complex structure and history ~\cite{McMillan:2016jtx}.

\par With the halo mass distribution established, we estimate the total stellar mass $M_s$ in local galaxies (at $z\approx 0$) with a halo mass of $M_{\text{h}}\equiv x M_0$, following the methodology in ~\cite{Behroozi:2019kql,Cui:2021hlu}:
\begin{equation}
    \log _{10}\left(\frac{M_s}{M_0}\right) = \epsilon_0-\log _{10}\left(10^{-\alpha_0 x}+10^{-\beta_0 x}\right)+\gamma_0 \exp^{-\frac{1}{2}\left(\frac{x}{\delta_0}\right)^2},
\end{equation}
where $M_0 =10^{12.06}M_\odot$, $\epsilon_0=-1.459$, $\alpha_0=1.972$, $\beta_0=0.488$, $\gamma_0=10^{-0.958}$, $\delta_0=0.391$.
\par In this paper, we adopt two typical structures for the stellar matter in galaxies: elliptical and disk.
\par We employ the Hernquist Model~\cite{Hernquist:1990be} to describe the isotropic distribution of elliptical galaxies:
\begin{equation}
    \rho_{s}^{e}(r_g) = \frac{C^e}{2\pi}\frac{R^e}{r_g(r_g+R^e)^3},
\end{equation}
where \(r_g\) represents the radial distance from the center of the halo, and \(R^e\), related to the half-mass radius, is defined as \(R^e = R_{1/2}/(1+\sqrt{2})\).
\par For disk galaxies, the stellar matter distribution is modeled using an exponential disk profile~\cite{2010gfe..book.....M,Cui:2021hlu}:
\begin{equation}
    \rho_s^d(R_g,h_g) = C^d\mathrm{exp}{\left(-\frac{R_g}{R^d}\right)}\mathrm{exp}{\left(-\frac{|h_g|}{h^d}\right)},
\end{equation}
where \((R_g, h_g)\) denote the cylindrical coordinates in the galaxy, with \(R^d\) and \(h^d\) being the characteristic scale lengths related to the galaxy's half-mass radius: \(R^d \approx R_{1/2}/1.68\) and \(h^d \approx R_{1/2}/10\).
\par In both cases, the half-mass radius of the halo is proportionate to the virial radius, expressed as \(R_{1/2} \approx 0.015 R_{\text{vir}}\)~\cite{Kravtsov:2012jn}. The normalization of the total stellar mass for both profiles is used to determine the normalization constants \(C^{e}\) and \(C^{d}\):
\begin{equation}
    M_s = \int \rho^{e(d)}_s(\vec{r}) d^3 \vec{r} ,
\end{equation}
where the integration is performed over the entire volume of the galaxy.\\

\section{Modification to the relative binary formation rate}\label{sec:Relative-Binary-Rate}
\par Here, we analyze the effects of black hole (BH) mass on the relative formation rates of BH-Neutron Star (NS) binaries, highlighting two primary factors.
\par Firstly, BHs of varying mass ranges exhibit distinct number densities. This variance stems from the progenitor stars' adherence to an Initial Mass Function (IMF)~\cite{Kroupa:2000iv}:
\begin{equation}
        \frac{\D n_{\text{star}}}{\D M_{\text{star}}}\propto M_{\text{star}}^{-2.3}, \ \ 0.5 M_{\odot} \leq M_{\text{star}} < 100 M_{\odot},
\end{equation}
assuming \(M_{\rm LMBH} = M_{\rm star}\) for lower mass black holes, which accrete the majority of their progenitor star's mass, and \(M_{\rm ABH} = \frac{1}{4}M_{\rm star}\) to account for mass loss during the evolution of more massive stars.
\par Secondly, for a BH-NS binary with a given initial orbital separation, the merger occurs more rapidly with increasing BH mass due to enhanced gravitational wave (GW) emission. This impacts the observed merger rates for GW detectors over observation periods of the order of 1 year. The orbital period's maximum value, allowing for a merger within \(\Delta t^{\rm ob} = 1\) year, is given by ~\cite{Maggiore:2007ulw}
\begin{equation}
    P^{\rm ob}_{\rm max}(M_{\rm BH}) \approx \left(\frac{256 \pi^{8/3} \Delta t^{\rm ob} G^{5/3} M_{\rm BH} M_{\rm NS}}{5 (M_{\rm BH} + M_{\rm NS})^{1/3}}\right)^{8/3},
\end{equation} 
with the period distribution for close binary systems modeled as a Gaussian centered at \(\log_{10}(P/\text{day}) = 4.8\) with a dispersion of 2.3~\cite{tokovinin2020formation}.
\par Integrating over the IMF and binary period distribution, we determine the modification to the merge rate ratio \(A\):
\begin{equation}
    A = \frac{\int_{3M_\odot}^{5M_\odot}\frac{\D n_{\text{star}}}{\D M_{\text{star}}}\D M_{\text{star}}\int^{{\rm log}P^{\rm ob}_{\rm max}(M_{\text{star}})}_{{\rm log}P^{\rm ob}_{\rm min}(M_{\text{star}})}\frac{\D n}{\D{\rm log}P}\D{\rm log}P}{\int_{20M_\odot}^{80M_\odot}\frac{\D n_{\text{star}}}{\D M_{\text{star}}}\D M_{\text{star}}\int^{{\rm log}P^{\rm ob}_{\rm max}(M_{\text{star}}/4)}_{{\rm log}P^{\rm ob}_{\rm min}(M_{\text{star}})}\frac{\D n}{\D{\rm log}P}\D{\rm log}P},
\end{equation}
yielding \(A=5.79\), under the assumption that \(M_{\rm NS} = 1.5M_\odot\) in all cases. This ratio highlights the significant impact of BH mass on BH-NS binary formation rates.

\acknowledgments

We thank Anil Seth, Dan Wik, Gail Zasowski and Zheng Zheng for useful discussions. This work is supported by the National Key Research and Development Program of China under Grant No. 2020YFC2201501.
S.G. is supported by the National Natural Science Foundation of China under Grant No. 12247147.
J.S. is supported by Peking University under startup Grant No. 7101302974 and the National Natural Science Foundation of China under Grants No. 12025507, No.12150015; and is supported by the Key Research Program of Frontier Science of the Chinese Academy of Sciences (CAS) under Grants No. ZDBS-LY-7003. Y.Z. is supported by the U.S. Department of Energy under Award No. DESC0009959.



\providecommand{\href}[2]{#2}\begingroup\raggedright\endgroup








\end{document}